\documentclass[runningheads]{llncs}

\usepackage{eccv}

\usepackage{eccvabbrv}

\usepackage{graphicx}
\usepackage{booktabs}

\usepackage[accsupp]{axessibility}  

\usepackage{hyperref}

\usepackage{orcidlink}

\newcommand{\tabincell}[2]{\begin{tabular}{@{}#1@{}}#2\end{tabular}}  
\usepackage{amsmath}
\usepackage{amssymb}
\usepackage{wrapfig}
\usepackage{float}
\usepackage{caption}
\usepackage{makecell}
\usepackage{enumitem}
\usepackage{multirow}
\usepackage{ dsfont }

\usepackage{comment}
\usepackage{bbm}
\usepackage{ulem}

\newcommand{\nickname}{BATseg}

\begin{document}

\title{\nickname{}: Boundary-aware Multiclass Spinal Cord Tumor Segmentation on 3D MRI Scans} 

\titlerunning{\nickname{}: Boundary-aware Multiclass Spinal Cord Tumor Segmentation ...}

\author{Hongkang Song\inst{1} \and
Zihui Zhang\inst{1} \and
Yanpeng Zhou\inst{2} \and
Jie Hu\inst{2} \and
Zishuo Wang\inst{2}\and
Hou Him Chan\inst{3} \and
Chon Lok Lei\inst{3} \and
Chen Xu\inst{4} \and
Yu Xin\inst{2} \and
Bo Yang\inst{1}} 

\authorrunning{H. Song et al.}

\institute{vLAR Group, The Hong Kong Polytechnic University \and
Department of Neurosurgery, Beijing Tiantan Hospital, Capital Medical University \and
Department of Biomedical Sciences, Faculty of Health Sciences, University of Macau \and
Insititute of Software Chinese Acadamy of Sciences}

\maketitle

\begin{abstract}
Spinal cord tumors significantly contribute to neurological morbidity and mortality. Precise morphometric quantification, encompassing the size, location, and type of such tumors, holds promise for optimizing treatment planning strategies. 
Although recent methods have demonstrated excellent performance in medical image segmentation, they primarily focus on discerning shapes with relatively large morphology such as brain tumors, ignoring the challenging problem of identifying spinal cord tumors which tend to have tiny sizes, diverse locations, and shapes. To tackle this hard problem of multiclass spinal cord tumor segmentation, we propose a new method, called \textbf{\nickname{}}, to learn a tumor surface distance field by applying our new multiclass boundary-aware loss function. To verify the effectiveness of our approach, we also introduce the first and large-scale spinal cord tumor dataset. It comprises gadolinium-enhanced T1-weighted 3D MRI scans from 653 patients and contains the four most common spinal cord tumor types: astrocytomas, ependymomas, hemangioblastomas, and spinal meningiomas.
Extensive experiments on our dataset and another public kidney tumor segmentation dataset show that our proposed method achieves superior performance for multiclass tumor segmentation. 

\end{abstract}

\section{Introduction}
\label{sec:intro}
\begin{figure*}[t]
\centering
\centerline{\includegraphics[width=\textwidth]{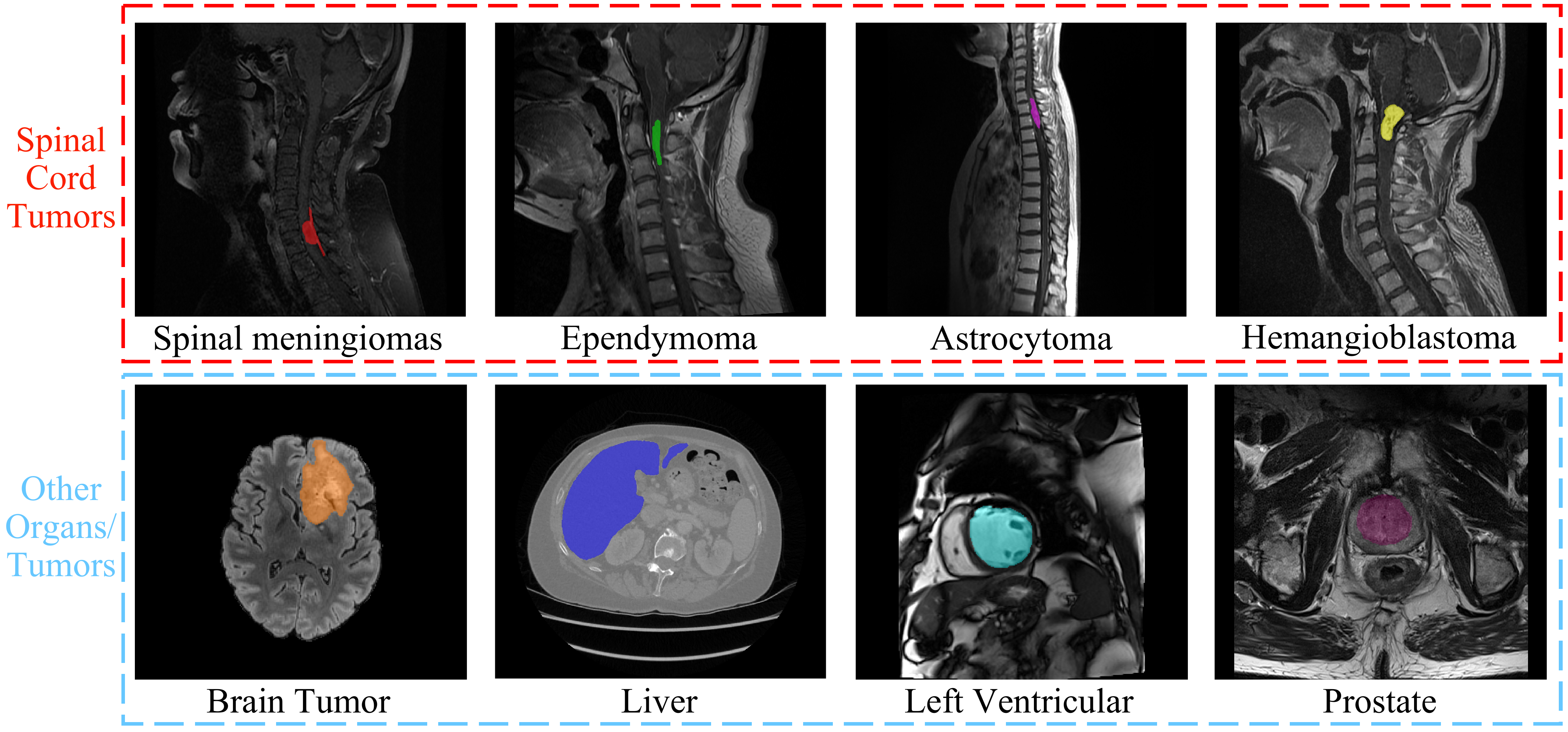}}
    \caption{An illustration of four types of spinal cord tumors shown in the \textit{first} row, and other commonly studied organs/tumors shown in the \textit{second} row.}
    \label{fig:differences}
\end{figure*}

The spinal cord, a pivotal central nervous system component, is critical in somatosensory perception and motor function\cite{kirshblum2018spinal}. Spinal cord tumors are primarily identified through 3D magnetic resonance imaging (MRI) scans. In the clinical arena, distinguishing between various types of spinal cord tumors remains a formidable challenge, leading to misdiagnoses, failures in tumor growth monitoring, and subsequent delays in therapeutic intervention. 

In past years, a series of sophisticated methods based on CNNs \cite{Ronneberger2015,Cicek2016,Isensee2020}, attention \cite{Hatamizadeh2022a,Lee2023}, and large models \cite{Ma2024,Liu2023} have been proposed, demonstrating excellent segmentation performance on a variety of medical images, thanks to the availability of large-scale datasets \cite{Simpson2019,Menze2015a}. However, there is still a lack of an automatic model that can precisely segment and recognize multiple types of spinal cord tumors. The main reasons are twofold.

First, existing methods predominantly focus on discerning shapes with relatively large morphology such as brain tumors, left/right ventricles, abdominal organs, \textit{etc.}, ignoring the challenging problem of spinal cord tumor identification. As illustrated in Figure \ref{fig:differences}, located in the long cylindrical spinal cord, common spinal cord tumors (\textit{colored pixels in the first row}) tend to have tiny sizes, diverse locations, and shape variations, whereas other tumors/organs (\textit{colored pixels in the second row}) exhibit relatively uniform and large dimensions. This means that a na\"ive application of existing methods to spinal cord tumor segmentation would result in inferior accuracy due to the stark anatomical disparities between spinal cord tumors and other body organs/tumors.

Second, there is a lack of public datasets containing multiple categories of spinal cord tumors to train deep neural networks. According to the recent MICCAI Grand Challenge \cite{GrandChallenge2024} and other competitions \cite{Simpson2019,Menze2015a}, the majority of public datasets target anatomical regions of the head and neck, abdomen, thorax, \textit{etc.}. The recent work \cite{reza2019cascaded} is among the early studies of spinal cord tumor segmentation, but its dataset has not been made public yet. To the best of our knowledge, there is no public dataset for spinal cord tumor segmentation.  

In this paper, we aim to tackle the challenging problem of multiclass spinal cord tumor segmentation. Firstly, we introduce a large-scale spinal cord tumor segmentation dataset meticulously curated to encompass gadolinium-enhanced T1-weighted 3D MRI scans from a cohort of 653 patients belonging to the four most prevalent tumor types. Secondly, we introduce a simple yet effective multiclass boundary-aware loss function that aids the backbone network such as nnUNet \cite{Isensee2020} to precisely segment multiple types of spinal cord tumors. 

\begin{figure*}[t]
\centering
\centerline{\includegraphics[width=\textwidth]{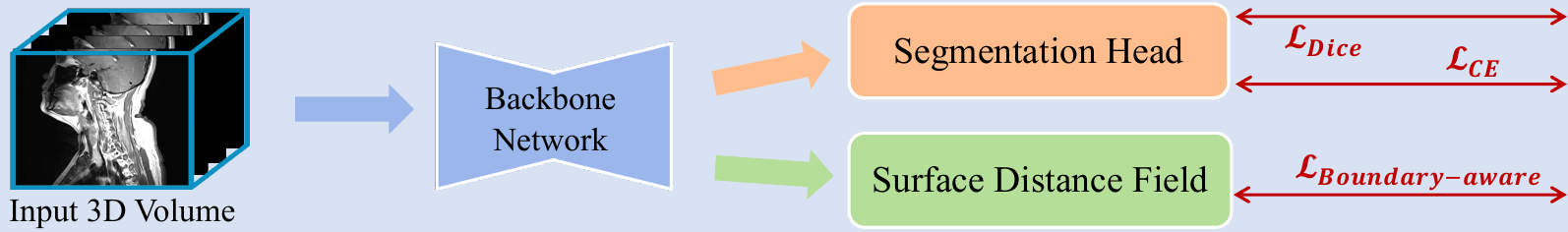}}
    \caption{An illustration of the overall framework.}
    \label{fig:pipeline_overview}
\end{figure*}

In particular, as illustrated in Figure \ref{fig:pipeline_overview}, to capture fine details of the tumor boundary regions, we add a new branch to an existing backbone network to learn a tumor surface distance field, in parallel to the segmentation head which predicts multiclass logits. Our boundary-aware loss, when jointly trained with the existing cross-entropy and Dice losses \cite{Isensee2020}, 
drives the new branch to learn a truncated normalized distance field for each tumor surface within a 3D MRI volume, thus helping the predicted tumor regions to be precisely tightened by a well-bounded 3D surface. By contrast, existing methods are usually only trained with cross-entropy and/or Dice losses on multiclass logits, often resulting in low probability predictions on tumor boundaries and therefore obtaining inferior segmentation performance. Furthermore, as the 3D surfaces of different types of spinal cord tumors could be vastly different, our boundary-aware loss is applied for each type of tumor separately, thus pushing the network to learn a more accurate surface distance field for each type of tumor. Our contributions are:
\begin{itemize}[leftmargin=*]
\setlength{\itemsep}{1pt}
\setlength{\parsep}{1pt}
\setlength{\parskip}{1pt}
    \item We present the first and large-scale dataset for spinal cord tumor segmentation, comprising four prevalent tumor types from 653 patients. 
    \item We introduce a \textbf{b}oundary-\textbf{a}ware loss to aid the network in learning 3D surfaces for each type of spinal cord tumor within 3D MRI volumes, capturing the fine details of tumor surrounding regions for accurate \textbf{t}umor \textbf{seg}mentation.
    \item We demonstrate superior 3D segmentation results on our multiclass spinal cord tumor dataset and another public dataset KNIGHT for Kidney tumor segmentation, showing better results than existing baselines.
\end{itemize}
We hope our method, named \textbf{\nickname{}}, and dataset could serve as a solid baseline for future research. Our data/code are available at \textcolor{red}{\href{https://github.com/vLAR-group/BATseg}{\tiny{https://github.com/vLAR-group/BATseg}}}

\section{Related Works}
\phantom{xx} \textbf{Medical Image Segmentation Methods:} Following the popularity of convolutional neural nets (CNNs), early methods mainly adopt \textbf{\textit{CNN-based frameworks}} for medical image segmentation, including U-Net \cite{Ronneberger2015}, 3D-UNet \cite{Cicek2016}, Y-Net \cite{B2018}, KiU-Net \cite{Valanarasu2020}, U-Net++ \cite{Zhou2020}, U-Net3+ \cite{Huang2020}, nnUNet \cite{Isensee2020}, and many other variants \cite{Wang2022f}. Among them, nnUNet \cite{Isensee2020} is a generalized segmentation framework which can configure the network architecture and settings automatically to extract features at multiple scales, showing particularly strong performance over various medical datasets. In this paper, we choose it as the backbone network and apply our boundary-aware loss function.

With the success of attention mechanism and vision transformers to capture long-range context information, many works extend \textbf{\textit{attention-based methods}} into CNN-based frameworks for better medical segmentation, including Trans U-Net \cite{Chen2021c}, UNETR \cite{Hatamizadeh2022a}, Swin UNETR \cite{Hatamizadeh2022}, nnFormer \cite{Zhou2023}, 3D UX-Net \cite{Lee2023}, and many other variants \cite{Wang2022f}. Very recently, large segmentation \textbf{\textit{foundation models}} have achieved tremendous advancements in natural images \cite{Kirillov2023}, and many succeeding works extend them to the field of medical image segmentation \cite{Ma2024,Liu2023,Wu2023,Wong2024}. Thanks to the powerful network architecture and large training datasets, these methods demonstrate excellent performance with the price of heavy computation resources and careful prompt engineering skills.      

\textbf{Spinal Cord Tumor Segmentation:}
Reza \textit{et al.}~\cite{reza2019cascaded} presented dual cascaded 3D CNNs for chordoma segmentation. However, the segmented tumors are located within the spinal region, and they display distinct intensities and dimensions and are juxtaposed with different tissue types. Consequently, such a segmentation model designed for the spine does not apply to our spinal cord tumor segmentation due to the inherent differences. Lemay \textit{et al.}~\cite{lemay2021automatic} use a two-level cascade architecture comprising two U-Net models to localize the spinal cord and segment the tumor separately, but it falls short in determining the multiple types of tumors, due to its weaknesses in discerning fine shapes. 

\textbf{Boundary-Aware Segmentation:} To improve segmentation performance on 2D natural images, prior works \cite{Zhao2019b, Qin2021, Borse2021,VuNgoc2021,Wang2021j} integrate boundary information into the learning process. However, they primarily focus on regressing edge pixels only, ignoring fine details of a wider context of shape surfaces. A few recent works \cite{Lin2019,Zhang2024,Bogensperger2024,Islam2024} leverage signed distances for medical image segmentation. However, they either focus on simplistic 2D cases or need complex deformation or transformation processes. Besides, they also fail to take into account boundary variations across multiclass. By comparison, our boundary-aware module is simple and learns accurate 3D surface distance fields for multiple classes.

\section{Spinal Cord Tumor Dataset}
\phantom{xxxx}\textbf{Data Collection:}
Our dataset encompasses anonymized 3D MRI scans of 653 patients from Beijing Tiantan Hospital, Capital Medical University under the approval of a national project, collected from October 2017 to September 2023, earmarked for the preoperative assessment of spinal cord tumors without prior treatment intervention. Ethical approval for this study was obtained from the Institutional Review Board of Beijing Tiantan Hospital, Capital Medical University (KY2022-114-02). All data are strictly limited to research purposes.

This dataset is distinguished by its comprehensive vertebral representations (encompassing cervical, thoracic, and lumbar regions) and includes the four predominant spinal cord tumor types: 1) \textit{spinal meningiomas} (n=247 patients), 2) \textit{ependymomas} (n=203), 3) \textit{astrocytomas} (n=101), and 4) \textit{hemangioblastomas} (n=102) as illustrated in the first row of Figure \ref{fig:differences}. 
Each patient's MRI volume is comprised of gadolinium-enhanced T1-weighted Sagittal (T1SC) MRI slices.
As summarized in Table \ref{tab:dataset_info}, the native resolution of the sagittal MRI volumes in this dataset varies, with in-plane resolutions ranging from 0.34 to 1.06 millimeters, and slice thicknesses spanning from 1.5 to 8 millimeters. The MRI volume with a varying number of slices in the sagittal plane, ranging from a minimum of 9 to a maximum of 36.

\textbf{Tumor Annotations:} Each patient's MRI volume is accompanied by expertly curated manual tumor annotations. The ground truth tumor areas were manually labeled by two independent neuroradiologists on the gadolinium-enhanced T1-weighted MRI scans. In cases where the demarcation of tumor areas was subject to dispute, the final decision was deferred to a senior neuroradiologist, ensuring the highest fidelity and precision in data annotation.

\textbf{Data Partitions:} For a more comprehensive and fairer comparison, our spinal cord tumor dataset is evenly partitioned into five distinct folds, leaving for a robust 5-fold cross evaluation of various methods and future studies. Details of the fivefold information are summarized in Table \ref{tab:dataset_folds}. 

\textbf{Data Pre-processing:}
Following the widely adopted standardization protocol in existing works \cite{Isensee2020,Zhou2023}, the resolution of sagittal slices is re-scaled to 0.47 mm for both the anterior-posterior and superior-inferior directions, and 3.3 mm for the right-left direction. The intensity values of MRI volumes are trilinearly interpolated and the tumor annotations are interpolated via nearest neighbors. Each volume's intensity is normalized by first subtracting the mean intensity value from each voxel and then dividing by the volume's standard deviation.

\begin{table}[t]
\caption{Details of MRI scans in our spinal cord tumor dataset.}\label{tab:dataset_info}
\tabcolsep= 0.1cm 
\centering
\resizebox{0.96\textwidth}{!}{
\begin{tabular}{l|c|c|c|c}
\toprule
 & \tabincell{c}{Spinal\\meningiomas}  & Ependymoma  & Astrocytoma & Hemangioblastoma\\
\hline
No. of subjects & 247   & 203 & 101  & 102 \\
Tumor size ($cm^{3}$) & 2.8$\pm$2.4  & 4.9$\pm$5.2 & 5.8$\pm$6.0  & 2.5$\pm$3.7 \\
In-plane resolutions ($mm$)&  0.601$\pm$0.124  & 0.516$\pm$0.128 & 0.579$\pm$0.161  & 0.558$\pm$0.155 \\
Slice thicknesses ($mm$)& 4.011$\pm$ 0.547  & 3.766$\pm$0.472 & 3.684$\pm$0.475  & 3.625$\pm$0.392 \\
No. of slices per subject &  11.0$\pm$0.206  & 11.0$\pm$0.438 & 11.4$\pm$2.501  & 11.1$\pm$0.421 \\
\bottomrule
\end{tabular}} 
\end{table}

\begin{table}[t]
\caption{The number of subjects in each of the five folds.}\label{tab:dataset_folds}
\tabcolsep= 0.1cm 
\centering
\resizebox{0.8\textwidth}{!}{
\begin{tabular}{l|c|c|c|c|c}
\toprule
 & \tabincell{c}{Spinal\\meningiomas}  & Ependymoma  & Astrocytoma & Hemangioblastoma & Total\\
\hline
Fold 1 & 54   & 42 & 19  & 17 & 132\\
Fold 2  & 50   & 27 &  26 & 26 & 129\\
Fold 3 & 42   & 44 & 25  & 19 & 130\\
Fold 4 &  58  & 38 & 16  & 19  & 131\\
Fold 5  &  43  & 52 & 15  & 21  & 131\\
\bottomrule
\end{tabular}} 
\end{table}

\section{Method}
\subsection{Overview}
\begin{figure*}[t]
\centering
\centerline{\includegraphics[width=\textwidth]{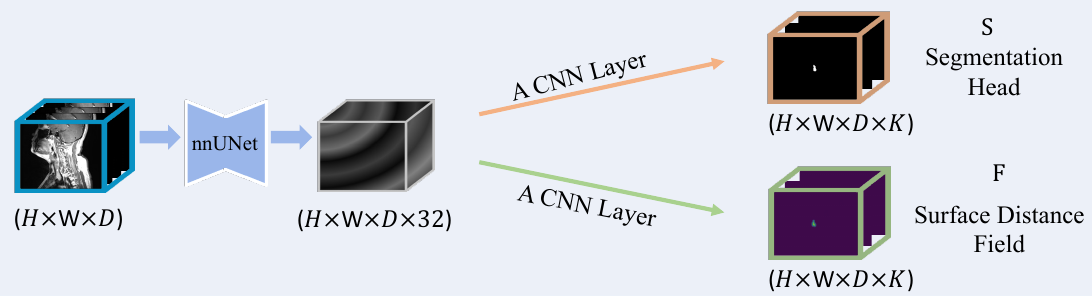}}
    \caption{The proposed segmentation pipeline. A 3D volume $T$ is fed into the backbone network, predicting per-voxel multiclass results $S$ via the segmentation head, and estimating the tumor surface distance field $F$ via a newly added head.}
    \label{fig:method}
\end{figure*}

As illustrated in Figure \ref{fig:method}, given an input 3D volume $T \in \mathds{R}^{H\times W \times D}$, where $H\times W$ denotes the slice resolution and $D$ represents the number of slices, we feed it into an existing backbone network such as nnUNet \cite{Isensee2020}, obtaining the per-voxel multiclass segmentation output $S \in \mathds{R}^{H\times W\times D\times K}$, where $K$ denotes the total number of classes. In this paper, $K$ is set as 5, representing the 4 types of spinal cord tumors together with the background class. 

In parallel to the segmentation head, we add another head to predict tumor surface distance field $F\in \mathds{R}^{H\times W\times D\times K}$, where a single convolutional layer is applied. The segmentation head is supervised by ground truth semantic labels using the common cross-entropy and Dice losses \cite{Isensee2020}. The surface distance field is supervised by ground truth labels using our multiclass boundary-aware loss. Details of neural layers are provided in Section \ref{sec:implementation}. We discuss the definition of tumor surface distance field in Section \ref{sec:surface_dis_field}, and the loss functions in Section \ref{sec:loss_functions}.  

\begin{figure*}[t]
\centering
\centerline{\includegraphics[width=\textwidth]{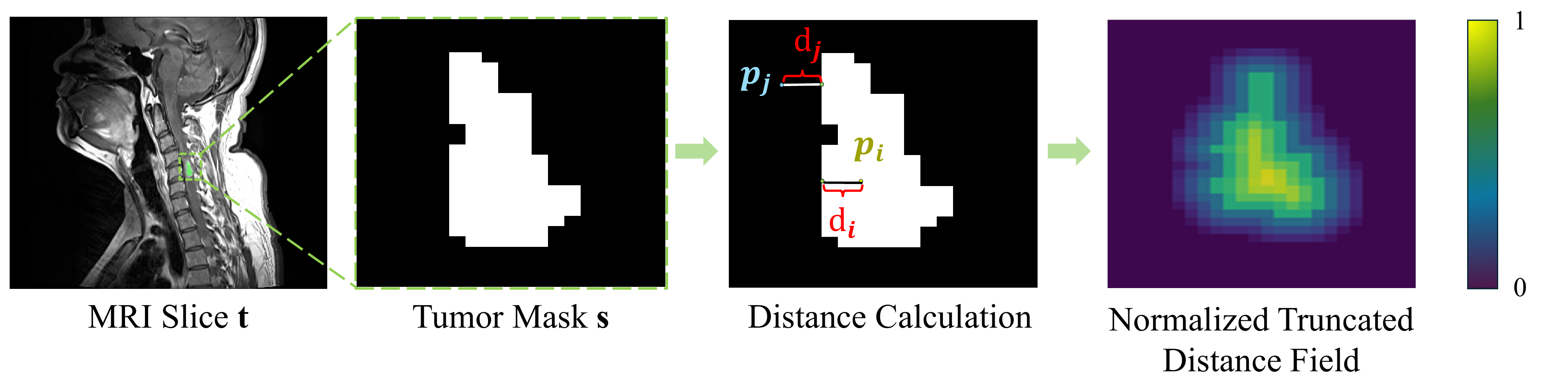}}
    \caption{An illustration of calculating tumor surface distance values on a single 2D slice.}
    \label{fig:Surface_distance_field}
\end{figure*}

\subsection{Tumor Surface Distance Field}\label{sec:surface_dis_field}

As illustrated in the first row of Figure \ref{fig:differences}, the spinal cord tumors exhibit particularly challenging shape variations. This motivates us to learn the complex and fine details of tumor boundaries and the surrounding regions in 3D volumes, by introducing the tumor surface distance field. 

As shown in Figure \ref{fig:Surface_distance_field}, for simplicity, we illustrate our definition of tumor surface distance field on a 2D slice, and our final definition on a 3D volume can be extended straightforwardly. Given a slice of MRI scan $\mathbf{t}\in \mathds{R}^{H\times W}$ belonging to a patient with any type of spinal cord tumor, and its corresponding ground truth tumor mask $\mathbf{s}\in \{0,1\}^{H\times W}$, where $1$ represents the foreground tumor pixel, $0$ the background. For any specific pixel $p_i$ within the tumor mask, we calculate its nearest distance (with a positive sign) to the tumor mask boundary, denoted as $d_i$. For any specific pixel $p_j$ outside the tumor mask, we also calculate its nearest distance (with a negative sign) to the tumor mask boundary, denoted as $d_j$. 

Since a spinal cord tumor usually only occupies a relatively small region, we opt to truncate the distance value to be zero for the pixel outside the tumor region, when its absolute distance $d_j$ is larger than the maximum distance values within the tumor mask. Formally it is defined as:
\begin{equation}\label{eq:truncation}
    d_j' \xleftarrow{truncated} d_j*\mathbbm{1}\Big(|d_j| \leq \max\{d_0 \cdots d_i \cdots d_I\}\Big)
\end{equation}
where $\{d_0 \cdots d_i \cdots d_I\}$ represents all distance values within the tumor mask, {$*$ is an element-wise multiplication, and $\mathbbm{1}()$ is an indicator function. Such a simple truncation allows the network to focus on the boundary surrounding details.  

Lastly, all non-truncated distance values are normalized within the range of $[0,1]$ as follows, while keeping all previously truncated values to be zeros. 
\begin{align}
    \bar{d}_j &= \Big(\frac{d_j'}{\max\{d_0 \cdots d_i \cdots d_I\}}+1\Big)/2 \nonumber  \\ \bar{d}_i &= \Big(\frac{d_i}{\max\{d_0 \cdots d_i \cdots d_I\}} +1 \Big)/2
\end{align}

Similarly, given a 3D volume $T$ and its ground truth tumor mask $S$, for every voxel inside the tumor mask, we calculate its surface distance value by measuring the nearest distance to the 3D tumor boundary, followed by truncation and normalization, obtaining the ground truth tumor surface distance values for the whole volume, denoted as $\bar{F}$, which will supervise our newly added network head. 

Note that, considering that each class of spinal cord tumors tends to have very different boundary shapes, the ground truth surface distance values are defined class-wise, instead of in a class-agnostic manner.  

\subsection{Loss Functions}\label{sec:loss_functions}

\phantom{xxx}\textbf{Cross-entropy and Dice Losses:} Following the existing work nnUNet \cite{Isensee2020}, the segmentation head is supervised by both cross-entropy and Dice losses, denoted as $\ell_{ce}$ and $\ell_{dice}$. Details can be found in \cite{Isensee2020}.  

\textbf{Boundary-Aware Loss:} Having the estimated tumor surface distance output $F \in \mathds{R}^{H\times W\times D \times K}$, and the corresponding ground truth distance label $\bar{F}$, a na\"ive method is to choose $\ell_1$ as the loss function to equally optimize every voxel. However, since spinal cord tumors only occupy relatively small regions, the majority of the surface distance values are zeros which belong to the truncated background voxels. To make our boundary-aware loss to focus on tumor boundary regions, we propose the following boundary-aware loss at each voxel:
\begin{equation}\label{eq:boundary_loss}
    \ell_{ba} = -\Big(f_{(h,w,d,k)} - \bar{f}_{(h,w,d,k)}\Big)^2*\Big|f_{(h,w,d,k)} - \bar{f}_{(h,w,d,k)}\Big|
\end{equation}
where $f_{(h,w,d,k)}$ is the predicted surface distance value, $\bar{f}_{(h,w,d,k)}$ is the ground truth at the $(h,w,d,k)^{th}$ voxel, and $*$ is an element-wise multiplication. The overall loss will be averaged across all voxels of the whole 3D volume. Intuitively, the majority of background voxels tend to be easily optimized to be zeros in the beginning, then our boundary-aware loss tends to focus on the voxels with non-zero distance values which belong to the interested boundary regions.  

Overall, the whole network is jointly trained with the three losses together as follows. For simplicity, we opt to equal weights for three terms.  
\begin{equation}
    \ell = \ell_{ce} + \ell_{dice} + \ell_{ba}
\end{equation}

\subsection{Implementation}\label{sec:implementation}
All experiments are conducted on a single NVIDIA 3090 GPU. Our model is trained for 1000 epochs with a learning rate of 0.01 with a decay of 0.00001 every epoch. 
Our backbone network nnUNet \cite{Isensee2020}, when trained on our spinal cord tumor dataset based on its automatically searched configurations, comprises 7 layers, each incorporating two convolutional operations. The initial two layers utilize a kernel size of $1\times1\times1$, while the subsequent layers employ a kernel size of $3\times3\times3$. The stride configuration for the first layer is $1\times1\times1$, for the fourth layer is $2\times2\times2$, and for the remaining layers it is $1\times2\times2$. The number of feature channels across all layers is set as 32, 64, 128, 256, 320, 320, and 320.

The backbone network, when trained on the public KNIGHT dataset \cite{Barros2023} based on its searched configurations, comprises six layers, each incorporating two $3\times3\times3$ convolutional operations. The stride configuration for the first layer is $1\times1\times1$, and for the remaining layers is $2\times2\times2$. The number of feature channels across all layers is set as 32, 64, 128, 256, 320, and 320.

\begin{figure*}[t]
\centering
\centerline{\includegraphics[width=\textwidth]{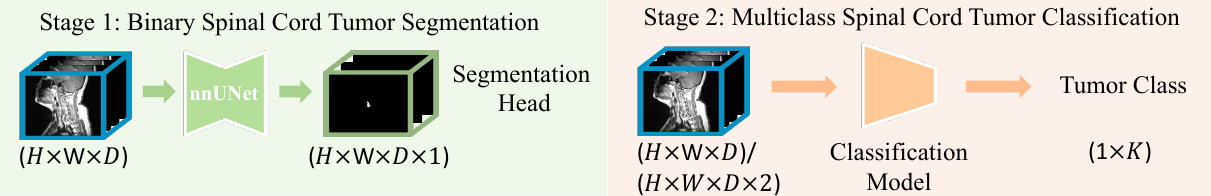}}
    \caption{The two-stage baseline for our problem of multiclass 3D segmentation. Given an input 3D volume, class-agnostic tumor voxels are segmented against the background in  Stage 1, followed by Stage 2 where the tumor type is classified for the same 3D volume. The predicted tumor type is then assigned to the estimated tumor mask.}
    \label{fig:two_stage_pipeline} 
\end{figure*}

\begin{table}[ht]
\caption{The Dice Coefficient (\%) and Hausdorff Distance (HD) (mm) scores over 5-fold cross validation on our spinal cord tumor dataset.
\textbf{Bold} numbers denote the best performance.}
\label{tab:quantitative_spinal_cord}
\resizebox{\textwidth}{!}{
\begin{tabular}{r|cc|cc|cc|cc|cc}
\toprule
\multicolumn{1}{c|}{\multirow{2}{*}{Methods}}& \multicolumn{2}{c|}{Spinal meningiomas} &\multicolumn{2}{c|}{Ependymoma} &\multicolumn{2}{c|}{Astrocytoma}&\multicolumn{2}{c|}{Hemangioblastoma}&\multicolumn{2}{c}{Mean}\\
\cline{2-11} 
\multicolumn{1}{c|}{}&\multicolumn{1}{c}{Dice \textuparrow}& \multicolumn{1}{c|}{HD \textdownarrow}
&\multicolumn{1}{c}{Dice \textuparrow}& \multicolumn{1}{c|}{HD \textdownarrow}
&\multicolumn{1}{c}{Dice \textuparrow}& \multicolumn{1}{c|}{HD \textdownarrow}
&\multicolumn{1}{c}{Dice \textuparrow}& \multicolumn{1}{c|}{HD \textdownarrow}
&\multicolumn{1}{c}{Dice \textuparrow}& \multicolumn{1}{c}{HD \textdownarrow}\\
\hline
nnFormer~\cite{Zhou2023} & 37.1 & 202.4& 0.0 &450.0&0.0&450.0&0.0&450.0&9.3&388.1\\
3D UX-Net~\cite{Lee2023} &40.3 &229.1&22.1&299.0&8.1&361.7&28.8&286.9&24.8&294.2\\
Swin UNETR~\cite{Hatamizadeh2022} &41.3& 220.7&20.7&305.0&7.3&370.2&26.7&301.4&24.0&299.3\\
nnUNet~\cite{Isensee2020} & 74.0 &69.1&43.5&195.3&26.0&270.1&67.7&111.0&52.8&161.4\\
\hline
\textbf{Two-Stage Methods} &&&&&&&&&&\\
\makecell[r]{\scriptsize{nnUNet+ResNet$^{with}_{mask}$}}& 33.4 &288.2&3.9&428.2&0.16&446.7&0.0&450.0&9.4&403.3\\
\makecell[r]{\scriptsize{nnUNet+ResNet$^{w/o}_{mask}$}} & 26.7 &318.0&7.5&408.7&0.0&450.0&0.0&450.0&8.6&406.7 \\
\makecell[r]{\scriptsize{nnUNet+UEnc$^{with}_{mask}$}}& 66.9 &128.1&37.7&241.8&28.1&303.8&57.6&168.8&47.6&210.6\\
\makecell[r]{\scriptsize{nnUNet+UEnc$^{w/o}_{mask}$}}& 53.7 &184.6&25.2&309.7&11.3&397.4&14.0&384.8&26.1&319.1\\

\hline
\textbf{\nickname{} (Ours)} & \textbf{80.9}&\textbf{34.4}  & \textbf{57.8}&\textbf{117.3} & \textbf{35.2}&\textbf{205.2}  & \textbf{74.6}&\textbf{76.1} & \textbf{62.1}&\textbf{108.3}\\
\hline
\end{tabular}}
\end{table}

\section{Experiments}

\textbf{Baselines:} We compare against the following two groups of baselines in medical image segmentation. All models are trained from scratch.
\begin{itemize}[leftmargin=*]
\setlength{\itemsep}{1pt}
\setlength{\parsep}{1pt}
\setlength{\parskip}{1pt}
    \item End-to-end training baselines, including the established state-of-the-art models nnFormer \cite{Zhou2023}, 3D UX-Net \cite{Lee2023}, Swin UNETR \cite{Hatamizadeh2022}, and nnUNet \cite{Isensee2020}.
    \item Two-stage training baselines. For our problem of per-voxel multiclass 3D segmentation, a na\"ive pipeline is to apply a per-voxel binary segmentation (tumor \textit{vs} background), followed by a per-volume classification model which takes the 3D volume with/without the predicted binary tumor mask as input, as illustrated in Figure \ref{fig:two_stage_pipeline}. For a fair comparison, in Stage 1, we choose the same backbone nnUNet \cite{Isensee2020} for binary segmentation. In Stage 2, we choose two powerful models: 3D ResNet101 \cite{He2016b} and the encoder part of nnUNet \cite{Isensee2020} (denoted as UEnc), both followed by  MLP layers with (1024-512-256-K) neurons. When training in Stage 2, we also feed the input 3D volume with or without the estimated class-agnostic tumor mask for comparisons. In total, we have 4 baselines: 1) nnUNet+ResNet$^{with}_{mask}$, 2) nnUNet+ResNet$^{w/o}_{mask}$, 3) nnUNet+UEnc$^{with}_{mask}$, 4) nnUNet+UEnc$^{w/o}_{mask}$. 
\end{itemize}

\textbf{Metrics:} Following M\&Ms Challenge\cite{Campello2021}, we use Dice Coefficient\cite{dice1945measures,sorensen1948method} and 95th percentile Hausdorff Distance\cite{rockafellar2009variational} as metrics. The Hausdorff Distance is assigned as a maximum value of 450 millimeters for missing predictions.

\textbf{Datasets:} In addition to evaluating all models on our spinal cord tumor dataset using 5-fold cross validation, we also evaluate all methods on the public KNIGHT dataset \cite{Barros2023} as the kidney tumors share similarity with spinal cord tumors. This dataset comprises 400 CT scans in 3D, which are partitioned into a training set of 300 volumes and a test set of 100 volumes. Each voxel is annotated by one of the two tumor classes: No Adjuvant Therapy (NoAT) and Candidate for Adjuvant Therapy (CanAT), or the background class. The resolution of all CT volumes is re-scaled to $2mm \times 2mm \times 2mm$. The intensities are trilinearly interpolated and the tumor annotations are interpolated via nearest neighbors. Each volume's intensities are normalized by first subtracting the mean intensity value from each voxel and then dividing by the volume's standard deviation.

\subsection{Results on Spinal Cord Tumor Dataset}
Table \ref{tab:quantitative_spinal_cord} compares the quantitative results of all baselines and our method on the spinal cord tumor dataset, averaged over 5-fold cross validation. Detailed quantitative results for each fold are provided in Appendix Tables 9/10/11/12/13.
Qualitative results are presented in Figure \ref{qualitative_results}. We can see that:
\begin{itemize}[leftmargin=*]
    \item Among the first group of baselines, nnUNet \cite{Isensee2020} demonstrates solid performance, achieving better Dice scores and Hausdorff Distances than other attention-based approaches.
    \item  Among the second group of baselines, the nnUNet encoder based classifiers yield higher Dice scores than ResNet101 based models.
    \item Compared with all these strong baselines, our method achieves significantly better performance in terms of both Dice scores and Hausdorff Distances. Notably, compared with nnUNet \cite{Isensee2020} which is our backbone network, our method obtains 10\% higher Dice score and 53.1mm better Hausdorff Distance, clearly demonstrating the effectiveness of our proposed tumor surface distance field learned by boundary-aware loss.
    \item We notice that the tumor type \textit{astrocytoma} (35.2\%) is significantly lower than others. A key reason is the similar appearance of \textit{astrocytoma} and \textit{ependymoma}, leading to frequent misclassification between these two tumor types. Consequently, the Dice score for \textit{ependymoma} is also relatively low (57.8\%).
\end{itemize}

\subsection{Results on KNIGHT Dataset}

\begin{table}[t]
\caption{The Dice Coefficient (\%) and Hausdorff Distance (HD) (mm) on the KNIGHT dataset \cite{Barros2023}.
\textbf{Bold} numbers denote the best performance.} 
\tabcolsep= 0.1cm 
\centering
\label{tab:quantitative_knight_2classes}
\resizebox{0.7\textwidth}{!}{
\begin{tabular}{r|cc|cc|cc}
\toprule
\multicolumn{1}{c|}{\multirow{2}{*}{Methods}}& \multicolumn{2}{c|}{NoAT} &\multicolumn{2}{c|}{CanAT} &\multicolumn{2}{c}{Mean}\\
\cline{2-7} 
\multicolumn{1}{c|}{}&\multicolumn{1}{c}{Dice \textuparrow}& \multicolumn{1}{c|}{HD \textdownarrow}
&\multicolumn{1}{c}{Dice \textuparrow}& \multicolumn{1}{c|}{HD \textdownarrow}
&\multicolumn{1}{c}{Dice \textuparrow}& \multicolumn{1}{c}{HD \textdownarrow}\\
\hline
nnFormer~\cite{Zhou2023} & 20.7 &238.3 & 13.5 &332.4&17.1&285.3\\

3D UX-Net~\cite{Lee2023} & 33.7 & 192.9& 14.5 &343.0&24.1&267.9\\

Swin UNETR~\cite{Hatamizadeh2022} & 31.9 & 211.4& 17.3 &328.7&24.6&270.0\\

nnUNet~\cite{Isensee2020} & 46.8 &\textbf{177.3} & 20.9 &320.2&33.8&248.7\\
\hline
\textbf{Two-Stage Methods} &&&&&&\\
\makecell[r]{nnUNet+ResNet$^{with}_{mask}$} & 44.4 &204.8&\textbf{34.0}&290.9&39.2&247.9\\
\makecell[r]{nnUNet+ResNet$^{w/o}_{mask}$} & 23.0 &330.3&22.6&337.0&22.8&333.6 \\
\makecell[r]{nnUNet+UEnc$^{with}_{mask}$}& \textbf{50.5 }&178.4&0.0&450.0&25.2&314.2 \\
\makecell[r]{nnUNet+UEnc$^{w/o}_{mask}$}& \textbf{50.5 }&178.4&0.0&450.0&25.2&314.2 \\
\hline
\textbf{\nickname{} (Ours)} & 48.0&181.9  & 33.4&\textbf{271.8} & \textbf{40.7}&\textbf{226.8}  \\
\hline
\end{tabular}} 
\end{table}

Table \ref{tab:quantitative_knight_2classes} compares the quantitative results of all methods on the KNIGHT dataset \cite{Barros2023}. Qualitative results are presented in Figure \ref{qualitative_results}. We can see that:
\begin{itemize}[leftmargin=*]
\setlength{\itemsep}{1pt}
\setlength{\parsep}{1pt}
\setlength{\parskip}{1pt}
\item Among the two groups of baselines, although nnUNet \cite{Isensee2020} and nnUNet+ ResNet$_{mask}^{with}$ show strong results, our method surpasses them all, achieving a Dice score of 40.7\% and a Hausdorff Distance of 226.8 mm. 

\begin{table}[H]
\caption{Ablation experiments in Group 1 on the KNIGHT dataset.
\textbf{Bold} numbers denote the best performance.}
\label{tab:ablation_distance_field}
\resizebox{\textwidth}{!}{
\begin{tabular}{r|cc|cc|cc}
\toprule
\multicolumn{1}{c|}{\multirow{2}{*}{Ablations}}& \multicolumn{2}{c|}{NoAT} &\multicolumn{2}{c|}{CanAT} &\multicolumn{2}{c}{Mean}\\
\cline{2-7} 
\multicolumn{1}{c|}{}&\multicolumn{1}{c}{Dice \textuparrow}& \multicolumn{1}{c|}{HD \textdownarrow}
&\multicolumn{1}{c}{Dice \textuparrow}& \multicolumn{1}{c|}{HD \textdownarrow}
&\multicolumn{1}{c}{Dice \textuparrow}& \multicolumn{1}{c}{HD \textdownarrow}\\
\hline
Distance Field (w/o truncation, w/o normalization, $\ell_{ba}$ based on $\ell_1$) &29.5&257.3&21.0&322.4&25.3&289.8\\
Distance Field (w/ truncation, w/o normalization, $\ell_{ba}$ based on $\ell_1$) &44.6&176.3&17.8&328.4&31.2&252.4\\
\textbf{(Ours) Distance Field} (w/ truncation, w/ normalization, $\ell_{ba}$ based on $\ell_1$)  &48.0&181.9&\textbf{33.4}&\textbf{271.8}&\textbf{40.7}&\textbf{226.8}\\
Distance Field (w/ truncation, w/ normalization, $\ell_{ba}$ based on $\ell_{2}$) &\textbf{50.6}&165.4&26.9&292.5&38.7&229.0
\\
Distance Field (w/ truncation, w/ normalization, $\ell_{ba}$ based on $\ell_{ce}$) &49.6&\textbf{162.7}&30.2&293.1&39.9&227.9\\
\hline
\end{tabular}} 
\end{table}

\begin{table}[H]
\caption{Ablation experiments in Group 2 on the KNIGHT dataset.
\textbf{Bold} numbers denote the best performance.}
\label{tab:ablation_distance_field_distance}
\tabcolsep= 0.1cm 
\centering
\resizebox{0.85\textwidth}{!}{
\begin{tabular}{r|cc|cc|cc}
\toprule
\multicolumn{1}{c|}{\multirow{2}{*}{Ablations}}& \multicolumn{2}{c|}{NoAT} &\multicolumn{2}{c|}{CanAT} &\multicolumn{2}{c}{Mean}\\
\cline{2-7} 
\multicolumn{1}{c|}{}&\multicolumn{1}{c}{Dice \textuparrow}& \multicolumn{1}{c|}{HD \textdownarrow}
&\multicolumn{1}{c}{Dice \textuparrow}& \multicolumn{1}{c|}{HD \textdownarrow}
&\multicolumn{1}{c}{Dice \textuparrow}& \multicolumn{1}{c}{HD \textdownarrow}\\
\hline
\textbf{(Ours) Truncated at $1*\max\{d_0 \cdots d_i \cdots d_I\}$} &48.0&181.9&\textbf{33.4}&271.8&\textbf{40.7}&226.8\\
Truncated at $2*\max\{d_0 \cdots d_i \cdots d_I\}$&47.7&173.2&25.4&312.5&36.5&242.8\\
Truncated at $3*\max\{d_0 \cdots d_i \cdots d_I\}$ &49.5&171.5&26.3&300.5&37.9&236.0  \\
\hline
\end{tabular}} 
\end{table}

\begin{table}[H]
\caption{Ablation experiments in Group 3 on the KNIGHT dataset.
\textbf{Bold} numbers denote the best performance.}
\label{tab:ablation_multiclass_binary}
\tabcolsep= 0.1cm 
\centering
\resizebox{0.8\textwidth}{!}{
\begin{tabular}{r|cc|cc|cc}
\toprule
\multicolumn{1}{c|}{\multirow{2}{*}{Ablations}}& \multicolumn{2}{c|}{NoAT} &\multicolumn{2}{c|}{CanAT} &\multicolumn{2}{c}{Mean}\\
\cline{2-7} 
\multicolumn{1}{c|}{}&\multicolumn{1}{c}{Dice \textuparrow}& \multicolumn{1}{c|}{HD \textdownarrow}
&\multicolumn{1}{c}{Dice \textuparrow}& \multicolumn{1}{c|}{HD \textdownarrow}
&\multicolumn{1}{c}{Dice \textuparrow}& \multicolumn{1}{c}{HD \textdownarrow}\\
\hline
Class-agnostic Distance Field &\textbf{48.6}&\textbf{174.7}&24.3&313.7&36.4&244.2\\
\textbf{(Ours) Multiclass Distance Field} &48.0&181.9&\textbf{33.4}&\textbf{271.8}&\textbf{40.7}&\textbf{226.8}\\
\hline
\end{tabular}}
\end{table}

\begin{table}[H]
\caption{Ablation experiments in Group 4 on the KNIGHT dataset.
\textbf{Bold} numbers denote the best performance.}
\label{tab:ablation_focal}
\tabcolsep= 0.1cm 
\centering
\resizebox{0.98\textwidth}{!}{
\begin{tabular}{r|cc|cc|cc}
\toprule
\multicolumn{1}{c|}{\multirow{2}{*}{Ablations}}& \multicolumn{2}{c|}{NoAT} &\multicolumn{2}{c|}{CanAT} &\multicolumn{2}{c}{Mean}\\
\cline{2-7} 
\multicolumn{1}{c|}{}&\multicolumn{1}{c}{Dice \textuparrow}& \multicolumn{1}{c|}{HD \textdownarrow}
&\multicolumn{1}{c}{Dice \textuparrow}& \multicolumn{1}{c|}{HD \textdownarrow}
&\multicolumn{1}{c}{Dice \textuparrow}& \multicolumn{1}{c}{HD \textdownarrow}\\
\hline
Boundary-aware loss: w/o $(f_{(h,w,d,k)} - \bar{f}_{(h,w,d,k)})^2$  &\textbf{48.9}&\textbf{164.1}&30.4&290.1&39.7&227.1\\

\textbf{(Ours) Boundary-aware loss}: w/ $(f_{(h,w,d,k)} - \bar{f}_{(h,w,d,k)})^2$ &48.0&181.9&\textbf{33.4}&\textbf{271.8}&\textbf{40.7}&\textbf{226.8}\\
Boundary-aware loss: w/ stop gradients $(f_{(h,w,d,k)} - \bar{f}_{(h,w,d,k)})^2$  &48.7&177.2&28.1&297.3&38.4&237.2
\\
\hline
\end{tabular}}
\end{table}

\begin{figure}[H]
\centering
\centerline{\includegraphics[width=\textwidth]{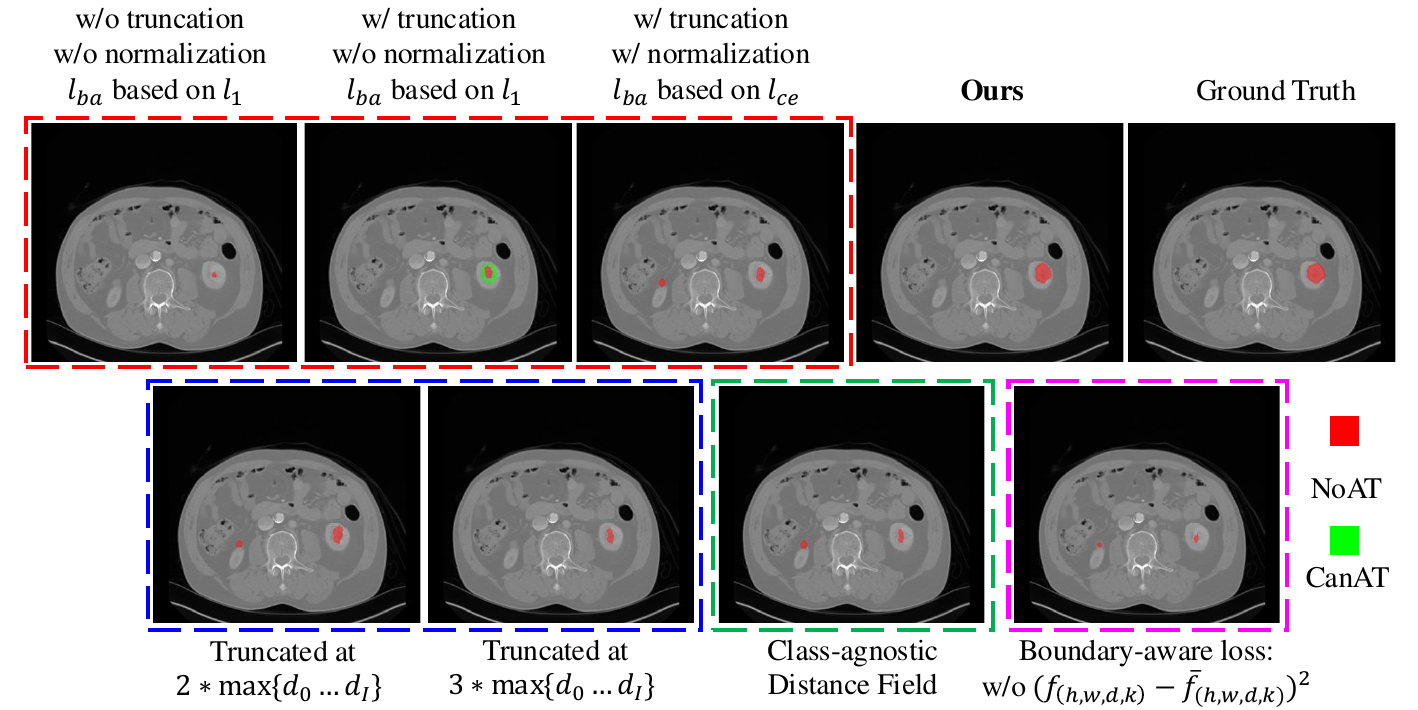}}
    \caption{Qualitative results of different ablation studies on the public KNIGHT dataset. Group 1/2/3/4 ablations are in the red, blue, green, and pink dotted boxes respectively.}
    \label{fig:ablation} 
\end{figure}

\item We notice that some two-stage methods (nnUNet+UEnc$^{with}_{mask}$ and nnUNet+ UEnc$^{w/o}_{mask}$) tend to classify all subjects into the NoAT category. As a consequence, they exhibit high performance in the NoAT class but yield zero Dice scores for the CanAT class.
\end{itemize}

On both datasets, as shown in Figure \ref{qualitative_results},
the tumor boundaries predicted by our method show greater consistency with the ground truth. In contrast, other baselines frequently predict larger tumors, leading to numerous false positive voxels. Additionally, some baselines tend to predict multiple sub-regions with different tumor types, whereas our method consistently generates more accurate masks with fine boundaries and correct tumor types.

\subsection{Ablation Study}

To validate the effectiveness of our design, we conduct the following groups of ablation studies on the public KNIGHT dataset. 
\begin{itemize}[leftmargin=*]
\setlength{\itemsep}{1pt}
\setlength{\parsep}{1pt}
\setlength{\parskip}{1pt}
\item Group 1: To verify our design of the tumor surface distance field, we choose four different settings: 1) the distance field is neither truncated nor normalized; 2) the distance field is truncated, but not normalized; 3) the distance field is truncated and normalized, which is the setting of our proposed method; 4) the $\ell_1$ term in our boundary-aware loss $\ell_{ba}$ is replaced by a $\ell_2$ loss; 5)the $\ell_1$ term in our boundary-aware loss $\ell_{ba}$ is replaced by a cross-entropy loss. 
\item Group 2: For the truncation strategy defined in Eq \ref{eq:truncation}, we verify three settings: the distances $d_j$ are truncated at one/two/three times of $\max\{d_0 \cdots d_i \cdots d_I\}$ respectively. Intuitively, the broader the truncation, the larger regions need to be focused, thus likely being less effective. 
\item Group 3: We further verify our design of the multiclass distance field. For comparison, we simplify the multiclass awareness to be a class-agnostic distance field. This means that the Surface Distance Field branch in Figure \ref{fig:method} has the output shape of $\mathds{R}^{H\times W\times D\times 1}$.
\item Group 4: For the boundary-aware loss defined in Eq \ref{eq:boundary_loss}, we verify the effectiveness of the term $(f_{(h,w,d,k)} - \bar{f}_{(h,w,d,k)})^2$. 
\end{itemize}

Tables \ref{tab:ablation_distance_field}/\ref{tab:ablation_distance_field_distance}/\ref{tab:ablation_multiclass_binary}/\ref{tab:ablation_focal} compare the ablation scores of Groups 1/2/3/4 respectively and Figure \ref{fig:ablation} show the qualitative results. We can see that:
\begin{itemize}[leftmargin=*]
\setlength{\itemsep}{1pt}
\setlength{\parsep}{1pt}
\setlength{\parskip}{1pt}
\item In Table \ref{tab:ablation_distance_field}, our design of truncation improves the Dice score by 5.9\%, and normalizing the truncated distance field further boosts the Dice score by an additional 9.5\%. In the meantime, the use of $\ell_{ba}$ based on $\ell_1$ is also better than $\ell_{ba}$ based on $\ell_{ce}$ or $\ell_{2}$.
\item In Table \ref{tab:ablation_distance_field_distance}, not surprisingly, the broader region to be truncated, the worse results obtained. The reason is that the fine details near the tumor boundary appear to be more important than the pixels far away from the boundary. 
\item In Table \ref{tab:ablation_multiclass_binary} our multiclass distance field simplifies the learning of complex boundaries, thus providing a clearer understanding of different tumor types, bringing a 4.3\% increase in the Dice score over the class-agnostic setting.
\item In Table \ref{tab:ablation_focal}, the term $(f_{(h,w,d,k)} - \bar{f}_{(h,w,d,k)})^2$ drives the model to focus more on the voxels with non-zero distance values, \textit{i.e.}, the interested tumor boundary region. This gives a 1\% improvement in the Dice score. Nevertheless, we emperically find that if we stop the gradients of $(f_{(h,w,d,k)} - \bar{f}_{(h,w,d,k)})^2$, the Dice score drops 2\%.
\end{itemize}

\begin{figure*}[t]
\centering
\centerline{\includegraphics[width=\textwidth]{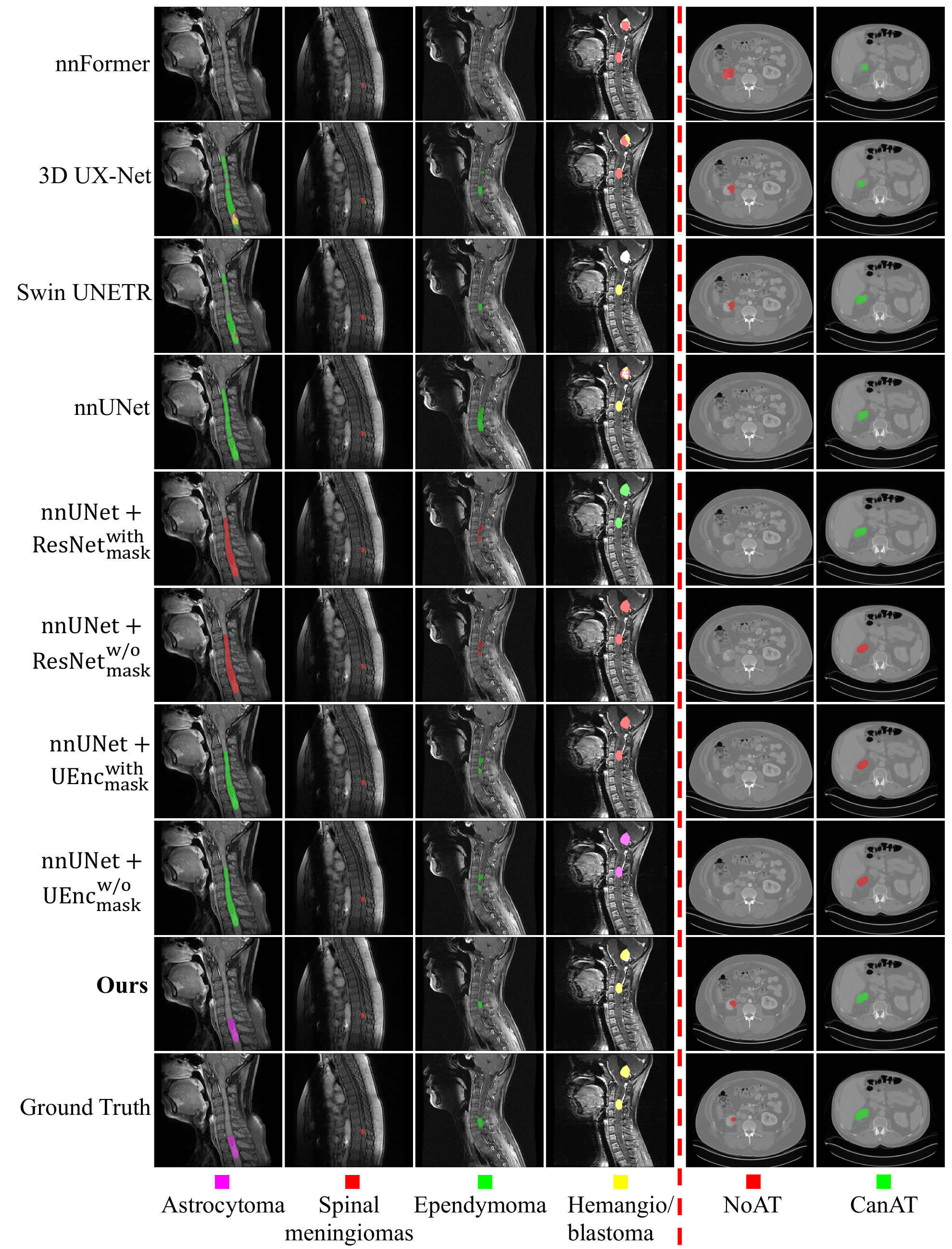}}
    \caption{Qualitative comparison of all methods on our spinal cord tumor dataset and the KNIGHT dataset.}
    \label{qualitative_results} 
\end{figure*}

\section{Conclusion}
In this paper, we present the first and large-scale multiclass spinal cord tumor segmentation dataset, meticulously compiled to include gadolinium-enhanced T1-weighted 3D MRI scans from a cohort of 653 patients. To capture the fine details of challenging spinal cord tumor boundary regions, we propose to learn a multiclass tumor surface distance field by applying a carefully designed boundary-aware loss function. 
Extensive experiments demonstrate that our method achieves superior segmentation accuracy, clearly outperforming existing baselines on both our spinal cord tumor dataset and a public kidney tumor segmentation dataset. One limitation of our study is that the generalizability of our model on datasets collected from different scanning equipment is yet to be explored, and we leave it for our future work.
We hope our new dataset and method could open up new opportunities in the field of spinal cord tumor segmentation. \\

\clearpage

\noindent\textbf{Acknowledgments:}
This work was supported by National Key R\&D Program of China under the Mainland-Hong Kong Joint Funding Scheme (Project ID from Innovation and Technology Fund in Hong Kong: MHP/012/21, Project ID from the Ministry of Science and Technology of China in Mainland: 2022YFE0201100),
and in part by the Innovation and Technology Commission of HKSAR under the Research Talent Hub for ITF projects (Pih/054/23, PiH/277/23, and InP/136/23), the University of Macau (reference no. SRG2024-00014-FHS, MYRG-CRG2022-00010-ICMS, and MYRG2022-00023-FHS), the Science and Technology Development Fund, Macao SAR (FDCT) (reference no. 0155/2023/RIA3, and 0048/2022/A).

%
%
\bibliographystyle{splncs04}
\bibliography{references}

 \clearpage

 \section{Appendix}
 \begin{table}[ht]
\caption{The Dice Coefficient (\%) and Hausdorff Distance (HD) (mm) scores on Fold 1 of our spinal cord tumor dataset.
\textbf{Bold} numbers denote the best performance.}

\label{tab2_fold1}
\resizebox{\textwidth}{!}{
\begin{tabular}{r|cc|cc|cc|cc|cc}
\toprule
\multicolumn{1}{c|}{\multirow{2}{*}{Methods}}& \multicolumn{2}{c|}{Spinal meningiomas} &\multicolumn{2}{c|}{Ependymoma} &\multicolumn{2}{c|}{Astrocytoma}&\multicolumn{2}{c|}{Hemangioblastoma}&\multicolumn{2}{c}{Mean}\\
\cline{2-11} 
\multicolumn{1}{c|}{}&\multicolumn{1}{c}{Dice \textuparrow}& \multicolumn{1}{c|}{HD \textdownarrow}
&\multicolumn{1}{c}{Dice \textuparrow}& \multicolumn{1}{c|}{HD \textdownarrow}
&\multicolumn{1}{c}{Dice \textuparrow}& \multicolumn{1}{c|}{HD \textdownarrow}
&\multicolumn{1}{c}{Dice \textuparrow}& \multicolumn{1}{c|}{HD \textdownarrow}
&\multicolumn{1}{c}{Dice \textuparrow}& \multicolumn{1}{c}{HD \textdownarrow}\\
\hline
nnFormer~\cite{Zhou2023} & 35.6 &230.1 &0.0  &450.0  &0.0  &450.0  &0.0  &450.0  & 8.9 & 395.0 \\

3D UX-Net~\cite{Lee2023}  & 37.3 & 253.0& 21.6 & 317.4 & 7.9 & 354.5 & 27.7 & 269.5 & 23.6 & 298.6 \\

Swin UNETR~\cite{Hatamizadeh2022}  & 48.8 & 167.6& 21.3 &303.2  &10.0  &363.3  &29.8  & 298.2 & 27.5 & 283.1 \\

nnUNet~\cite{Isensee2020}  & 73.4 & 56.2& 46.4 & 167.4 & 21.9 & 299.5 & 51.9 & 189.3 & 48.4 & 178.1 \\
\hline
\textbf{Two-Stage Methods}  &&&&&&&&&&\\
\makecell[r]{\scriptsize{nnUNet+ResNet$^{with}_{mask}$}} & 34.0 & 280.0&  0.0  &450.0  &0.0  &450.0  &0.0  &450.0 & 8.5 & 407.5 \\
\makecell[r]{\scriptsize{nnUNet+ResNet$^{w/o}_{mask}$}}  & 34.3 &271.9 & 0.0  &450.0  &0.0  &450.0  &0.0  &450.0  & 8.6 & 405.5 \\

\makecell[r]{\scriptsize{nnUNet+UEnc$^{with}_{mask}$}} & 62.3 & 140.7& 41.5 & 215.4 & \textbf{35.6} & 287.1 & 54.4 & 189.6 & 48.5 & 208.2 \\

\makecell[r]{\scriptsize{nnUNet+UEnc$^{w/o}_{mask}$}} &47.9  &203.0 & 25.5 & 304.2 &17.0  &381.5  & 6.5 & 416.8 & 24.2 & 326.4 \\

\hline
\textbf{\nickname{} (Ours)} &\textbf{78.4}&\textbf{40.4}&\textbf{54.7}&\textbf{151.2}&30.6&\textbf{206.6}&\textbf{76.3}&\textbf{80.6}&\textbf{60.0}&\textbf{119.7} \\
\hline
\end{tabular}} 
\end{table}

\begin{table}[!ht]
\caption{The Dice Coefficient (\%) and Hausdorff Distance (HD) (mm) scores on Fold 2 of our spinal cord tumor dataset.
\textbf{Bold} numbers denote the best performance.}
\label{tab2_fold2}
\resizebox{\textwidth}{!}{
\begin{tabular}{r|cc|cc|cc|cc|cc}
\toprule
\multicolumn{1}{c|}{\multirow{2}{*}{Methods}}& \multicolumn{2}{c|}{Spinal meningiomas} &\multicolumn{2}{c|}{Ependymoma} &\multicolumn{2}{c|}{Astrocytoma}&\multicolumn{2}{c|}{Hemangioblastoma}&\multicolumn{2}{c}{Mean}\\
\cline{2-11} 
\multicolumn{1}{c|}{}&\multicolumn{1}{c}{Dice \textuparrow}& \multicolumn{1}{c|}{HD \textdownarrow}
&\multicolumn{1}{c}{Dice \textuparrow}& \multicolumn{1}{c|}{HD \textdownarrow}
&\multicolumn{1}{c}{Dice \textuparrow}& \multicolumn{1}{c|}{HD \textdownarrow}
&\multicolumn{1}{c}{Dice \textuparrow}& \multicolumn{1}{c|}{HD \textdownarrow}
&\multicolumn{1}{c}{Dice \textuparrow}& \multicolumn{1}{c}{HD \textdownarrow}\\
\hline

nnFormer~\cite{Zhou2023} & 37.2 &193.3 & 0.0  &450.0  &0.0  &450.0  &0.0  &450.0  & 9.3 & 385.8 \\

3D UX-Net~\cite{Lee2023}  & 41.4 & 236.7& 14.1 & 350.8 & 9.0 & 348.8 & 36.9 & 241.0 & 25.4 & 294.3 \\

Swin UNETR~\cite{Hatamizadeh2022}  & 39.0 &236.6 & 12.2 & 347.0 & 10.6 & 366.7 & 24.7 & 310.2 & 21.6 & 315.1 \\

nnUNet~\cite{Isensee2020}  & 72.9 & 83.7& 33.0 & 255.2 & 32.6 & 206.0 & 64.8 & 119.2 & 50.8 & 166.0 \\
\hline
\textbf{Two-Stage Methods} &&&&&&&&&&\\
\makecell[r]{\scriptsize{nnUNet+ResNet$^{with}_{mask}$}} & 33.9 & 280.1& 0.0 & 450.0 & 0.0 & 450.0 & 0.0 & 450.0 & 8.5 & 407.5 \\
\makecell[r]{\scriptsize{nnUNet+ResNet$^{w/o}_{mask}$}}  & 33.3 &288.4 &0.0  &450.0  &0.0  & 450.0 & 0.0. & 450.0 & 8.3 & 409.6 \\

\makecell[r]{\scriptsize{nnUNet+UEnc$^{with}_{mask}$}} & 72.1 & 93.4& 21.3 & 317.5 & 33.0 & 258.0 & 69.8 & 123.2 & 49.1 & 198.0 \\
\makecell[r]{\scriptsize{nnUNet+UEnc$^{w/o}_{mask}$}} & 62.9 & 146.5& 22.7 & 328.3 & 24.0 & 319.8 & 3.5 & 432.7 & 28.3 & 306.8 \\

\hline
\textbf{\nickname{} (Ours)} & \textbf{82.2}&\textbf{35.0}  & \textbf{59.5}&\textbf{92.6} & \textbf{41.8}&\textbf{153.7}  & \textbf{72.7}&\textbf{77.3} & \textbf{64.1}&\textbf{89.7}\\
\hline
\end{tabular}} 
\end{table}

\begin{table}[!ht]
\caption{The Dice Coefficient (\%) and Hausdorff Distance (HD) (mm) scores on Fold 3 of our spinal cord tumor dataset.
\textbf{Bold} numbers denote the best performance.}
\label{tab2_fold3}
\resizebox{\textwidth}{!}{
\begin{tabular}{r|cc|cc|cc|cc|cc}
\toprule
\multicolumn{1}{c|}{\multirow{2}{*}{Methods}}& \multicolumn{2}{c|}{Spinal meningiomas} &\multicolumn{2}{c|}{Ependymoma} &\multicolumn{2}{c|}{Astrocytoma}&\multicolumn{2}{c|}{Hemangioblastoma}&\multicolumn{2}{c}{Mean}\\
\cline{2-11} 
\multicolumn{1}{c|}{}&\multicolumn{1}{c}{Dice \textuparrow}& \multicolumn{1}{c|}{HD \textdownarrow}
&\multicolumn{1}{c}{Dice \textuparrow}& \multicolumn{1}{c|}{HD \textdownarrow}
&\multicolumn{1}{c}{Dice \textuparrow}& \multicolumn{1}{c|}{HD \textdownarrow}
&\multicolumn{1}{c}{Dice \textuparrow}& \multicolumn{1}{c|}{HD \textdownarrow}
&\multicolumn{1}{c}{Dice \textuparrow}& \multicolumn{1}{c}{HD \textdownarrow}\\
\hline

nnFormer~\cite{Zhou2023} & 22.3 &274.0 &0.0  &450.0  &0.0  &450.0  &0.0  &450.0  &5.6  & 406.0 \\

3D UX-Net~\cite{Lee2023}  & 42.4 &222.2 & 23.6 & 287.0 & 10.3 & 341.1 & 30.6 & 266.4 & 26.7 & 279.2 \\

Swin UNETR~\cite{Hatamizadeh2022}  & 44.3 & 207.3& 24.2 & 268.0 & 8.2 & 357.1 & 27.7 & 292.9 & 26.1 & 281.3 \\

nnUNet~\cite{Isensee2020}  & 78.9 &60.1 & 47.1 & 156.2 & 35.1 & 212.3 & \textbf{80.7} & \textbf{34.9} & 60.5 & 115.9 \\
\hline
\textbf{Two-Stage Methods} &&&&&&&&&&\\
\makecell[r]{\scriptsize{nnUNet+ResNet$^{with}_{mask}$}} & 29.0 &313.3 & 3.6 & 430.7 & 0.0  &450.0  &0.0  &450.0  &8.2  & 411.0 \\
\makecell[r]{\scriptsize{nnUNet+ResNet$^{w/o}_{mask}$}}  & 25.6 & 329.8& 12.3 & 383.4 &  0.0  &450.0  &0.0  &450.0  & 9.5 & 403.3 \\

\makecell[r]{\scriptsize{nnUNet+UEnc$^{with}_{mask}$}} & 68.7 & 120.0& 40.7 & 228.9 & 15.0 & 366.4 & 39.5 & 248.6 & 41.0 & 241.0 \\
\makecell[r]{\scriptsize{nnUNet+UEnc$^{w/o}_{mask}$}} & 60.7 & 164.8& 25.7 & 306.3 & 0.0 & 450.0 & 27.8 & 322.6 & 28.6 & 310.9 \\

\hline
\textbf{\nickname{} (Ours)} & \textbf{86.2}&\textbf{23.1}  & \textbf{55.2}&\textbf{122.1} & \textbf{43.7}&\textbf{161.5}  & 76.9&37.0 & \textbf{65.5}&\textbf{85.9}\\
\hline
\end{tabular}}
\end{table}

\begin{table}[!ht]
\caption{The Dice Coefficient (\%) and Hausdorff Distance (HD) (mm) scores on Fold 4 of our spinal cord tumor dataset.
\textbf{Bold} numbers denote the best performance.}
\label{tab2_fold4}
\resizebox{\textwidth}{!}{
\begin{tabular}{r|cc|cc|cc|cc|cc}
\toprule
\multicolumn{1}{c|}{\multirow{2}{*}{Methods}}& \multicolumn{2}{c|}{Spinal meningiomas} &\multicolumn{2}{c|}{Ependymoma} &\multicolumn{2}{c|}{Astrocytoma}&\multicolumn{2}{c|}{Hemangioblastoma}&\multicolumn{2}{c}{Mean}\\
\cline{2-11} 
\multicolumn{1}{c|}{}&\multicolumn{1}{c}{Dice \textuparrow}& \multicolumn{1}{c|}{HD \textdownarrow}
&\multicolumn{1}{c}{Dice \textuparrow}& \multicolumn{1}{c|}{HD \textdownarrow}
&\multicolumn{1}{c}{Dice \textuparrow}& \multicolumn{1}{c|}{HD \textdownarrow}
&\multicolumn{1}{c}{Dice \textuparrow}& \multicolumn{1}{c|}{HD \textdownarrow}
&\multicolumn{1}{c}{Dice \textuparrow}& \multicolumn{1}{c}{HD \textdownarrow}\\
\hline

nnFormer~\cite{Zhou2023} & 53.3 & 107.4& 0.0 &450.0  &0.0  &450.0  &0.0  &450.0  &13.3  &364.4  \\

3D UX-Net~\cite{Lee2023}  & 43.3 & 192.7& 21.1 & 297.3 & 6.9 & 387.3 & 20.2 & 345.5 & 22.9 & 305.7 \\

Swin UNETR~\cite{Hatamizadeh2022}  &36.6  &255.3 & 16.9 & 347.2 & 2.2 & 404.8 & 15.5 & 363.1 & 17.8 & 342.6 \\

nnUNet~\cite{Isensee2020}  & 71.0 &78.7 & 40.7 & 214.1 & 23.3 & 310.7 & 70.5 & 85.5 & 51.4 & 172.3 \\
\hline
\textbf{Two-Stage Methods} &&&&&&&&&&\\
\makecell[r]{\scriptsize{nnUNet+ResNet$^{with}_{mask}$}} & 36.4 & 283.0& 0.0 & 450.0 & 0.0 & 450.0 & 0.0 & 450.0 & 9.1 & 408.3 \\
\makecell[r]{\scriptsize{nnUNet+ResNet$^{w/o}_{mask}$} } & 12.8 &386.4 & 23.3 & 322.7 & 0.0 & 450.0 & 0.0 & 450.0 & 9.0 & 402.3 \\

\makecell[r]{\scriptsize{nnUNet+UEnc$^{with}_{mask}$}} & 62.4 & 171.3&  42.5& 228.9 & \textbf{34.0} & 289.6 & 54.1 & 167.2 & 48.3 & 214.3 \\
\makecell[r]{\scriptsize{nnUNet+UEnc$^{w/o}_{mask}$}} &48.7  &203.3 & 18.1 & 345.9 & 15.7 & 385.5 & 22.8 & 344.0 & 26.3 & 319.7 \\

\hline
\textbf{\nickname{} (Ours)} & \textbf{79.0}&\textbf{39.4}  & \textbf{53.1}&\textbf{139.0} & 31.4&\textbf{258.0}  & \textbf{76.8}&\textbf{74.6} & \textbf{60.1}&\textbf{127.7}\\
\hline
\end{tabular}}
\end{table}

\begin{table}[!ht]
\caption{The Dice Coefficient (\%) and Hausdorff Distance (HD) (mm) scores on Fold 5 of our spinal cord tumor dataset.
\textbf{Bold} numbers denote the best performance.}
\label{tab2_fold5}
\resizebox{\textwidth}{!}{
\begin{tabular}{r|cc|cc|cc|cc|cc}
\toprule
\multicolumn{1}{c|}{\multirow{2}{*}{Methods}}& \multicolumn{2}{c|}{Spinal meningiomas} &\multicolumn{2}{c|}{Ependymoma} &\multicolumn{2}{c|}{Astrocytoma}&\multicolumn{2}{c|}{Hemangioblastoma}&\multicolumn{2}{c}{Mean}\\
\cline{2-11} 
\multicolumn{1}{c|}{}&\multicolumn{1}{c}{Dice \textuparrow}& \multicolumn{1}{c|}{HD \textdownarrow}
&\multicolumn{1}{c}{Dice \textuparrow}& \multicolumn{1}{c|}{HD \textdownarrow}
&\multicolumn{1}{c}{Dice \textuparrow}& \multicolumn{1}{c|}{HD \textdownarrow}
&\multicolumn{1}{c}{Dice \textuparrow}& \multicolumn{1}{c|}{HD \textdownarrow}
&\multicolumn{1}{c}{Dice \textuparrow}& \multicolumn{1}{c}{HD \textdownarrow}\\
\hline

nnFormer~\cite{Zhou2023} & 37.0 & 207.3& 0.0 &450.0  &0.0  &450.0  &0.0  &450.0  &9.3  & 389.3 \\

3D UX-Net~\cite{Lee2023}  & 36.9 & 240.7& 30.3 & 242.6 & 6.6 & 376.9 & 28.8 & 312.1 & 25.7 & 293.1 \\

Swin UNETR~\cite{Hatamizadeh2022}  & 37.6 & 236.7& 29.1 & 259.4 & 5.6 & 359.2 & 35.8 & 242.7 & 27.0 & 274.5 \\

nnUNet~\cite{Isensee2020}  & 73.6 & 66.9& 50.3 & 183.4 & 17.3 & 321.9 & \textbf{70.7} & 126.1 & 53.0 & 174.6 \\
\hline
\textbf{Two-Stage Methods} &&&&&&&&&&\\
\makecell[r]{\scriptsize{nnUNet+ResNet$^{with}_{mask}$}} & 33.9 & 284.8&  15.9& 360.3 & 0.8 & 433.3 & 0.0 & 450.0 & 12.7 & 382.1 \\
\makecell[r]{\scriptsize{nnUNet+ResNet$^{w/o}_{mask}$}}  & 27.5 & 313.7& 2.1 & 437.4 & 0.0 & 450.0 & 0.0 & 450.0 & 7.4 & 412.8 \\

\makecell[r]{\scriptsize{nnUNet+UEnc$^{with}_{mask}$}} & 68.9 & 115.1&  42.6&  218.6& 22.8 & 318.1 & 70.4 & 115.3 & 51.2 & 191.7 \\
\makecell[r]{\scriptsize{nnUNet+UEnc$^{w/o}_{mask}$}} & 48.5 & 205.3& 34.0 & 263.6 & 0.0 & 450.0 & 9.2 & 407.7 & 22.9 & 331.7 \\

\hline
\textbf{\nickname{} (Ours)} & \textbf{78.8}&\textbf{34.1}  & \textbf{66.7}&\textbf{81.4} & \textbf{28.4}&\textbf{246.2}  & 70.3&\textbf{111.2} & \textbf{61.1}&\textbf{118.2}\\
\hline
\end{tabular}}
\end{table}

\end{document}